# Advancing MG Energy Management: A Rolling Horizon Optimization Framework for Three-Phase Unbalanced Networks Integrating Convex Formulations.

Pablo Cortés, Alejandra Tabares, Jhon Franco

**Abstract**

Real-world three-phase microgrids face two interconnected challenges: (1) time-varying uncertainty from renewable generation and demand, and (2) persistent phase imbalances caused by uneven distributed energy resources (DERs), load asymmetries, and grid faults. Conventional energy management systems fail to address these challenges holistically and static optimization methods (e.g., day-ahead scheduling) lack adaptability to real-time fluctuations, while balanced three-phase models ignore critical asymmetries that degrade voltage stability and efficiency. This work introduces a dynamic rolling horizon optimization framework specifically designed for unbalanced three-phase microgrids. Unlike traditional two-stage stochastic approaches that fix decisions for the entire horizon, the rolling horizon algorithm iteratively updates decisions in response to real-time data (e.g., updated solar forecasts, load measurements). By solving a sequence of shorter optimization windows—each incorporating the latest system state and forecasts—the method achieves three key advantages: Adaptive Uncertainty Handling by continuously "re-plans" operations to mitigate forecast errors (e.g., sudden cloud cover reducing solar output). Phase Imbalance Correction by dynamically adjusts power flows across phases to minimize voltage deviations and losses caused by asymmetries, and computational Tractability, i.e., shorter optimization windows, combined with the mathematical mhodel, enable better decision-making holding accuracy. For comparison purposes, we derive three optimization models: a nonlinear nonconvex model for high-fidelity offline planning, a convex quadratic approximation for day-ahead scheduling, and a linearized model to important for theoretical reasons such as decomposition algorithms. Simulations on a 34-bus unbalanced microgrid demonstrate that the rolling horizon strategy reduces expected operational costs by 17% compared to two-stage stochastic methods. Crucially, the quadratic model achieves solutions in under 2 seconds per optimization window, critical for real-time deployment. Furthermore, the linear model achieves a similar performance, maintaining a 99.28% of the quadratic convex model's accuracy.

Nomenclature

*Sets*

| | |
|---|---|
| $\Omega$ | Set of buses with power demand |
| $BS$ | Set of batteries |
| $DG$ | Set of diesel generators |
| $GR$ | Subset of substations |
| $H$ | Set of phases |
| $L$ | Set of lines |
| $N$ | Set of buses |
| $PV$ | Set of photovoltaic power plants |
| $WT$ | Set of wind turbines |
| $T$ | Set of time periods inside the control window |

*Indices*

| | |
|---|---|
| $bs$ | Index of battery |
| $dg$ | Index of diesel generator |
| $f$ | Index of phase |
| $gr$ | Index of substation |
| $k, m, n$ | Indices of buses |
| $km, mn$ | Indices of lines |
| $pv$ | Index of photovoltaic power plants |
| $t$ | Index of time |
| $wt$ | Index of wind turbines |

*Parameters*

| | |
|---|---|
| $\beta_{bs}^{BS}$ | Self-discharge rate of batteries |
| $\beta$ | Discount factor |
| $\Delta t$ | Time duration of each time step |
| $\eta_{bs}^{BS}$ | Round-trip efficiency of battery |
| $B_{mn,f}$ | Shunt susceptance of line $mn$ in phase $f$ |
| $C_t^{dg}$ | Energy cost in period $t$ when produced with diesel generators |
| $C_t^{Gr}$ | Energy cost in period $t$ at substation |
| $\overline{E}_m^{BS}, \underline{E}_m^{BS}$ | Capacity limits of battery systems |
| $\overline{P}_m^{BS+}, \overline{P}_m^{BS-}$ | Maximum allowed power charge and discharge to battery located at bus $m$ |
| $pf$ | Power factor |
| $\overline{Q}_{dg}^{DG}, \overline{Q}_{pv}^{PV}, \overline{Q}_{wt}^{WT}$ | Maximum reactive power generation for DERs |
| $\underline{Q}_{dg}^{DG}, \underline{Q}_{pv}^{PV}, \underline{Q}_{wt}^{WT}$ | Minimum reactive power generation for DERs |
| $R_{mn,h}$ | Resistance of line $mn$ in phase $f$ |
| $X_{mn,h}$ | Reactance of line $mn$ in phase $f$ |
| $\overline{I}_{mn}$ | Current limit trough line $mn$ |
| $\overline{S}_{dg}^{DG}, \overline{S}_{pv}^{PV}, \overline{S}_{wt}^{WT}$ | Apparent power generation limit of DERs |
| $\overline{V}, \underline{V}$ | Voltage upper and lower limits |

*Variables*

| | |
|---|---|
| $E_{bs,t}^{BS}$ | Energy stored in battery at period $t$ |
| $I_{mn,f,t}^{Im}, I_{mn,f,t}^{Re}$ | Imaginary and real component of electrical current flowing through line $mn$, phase $f$ at period $t$ |
| $I_{bs,f,t}^{BS,Im}, I_{bs,f,t}^{BS,Re}$ | Imaginary and real component of current injected by battery at period $t$ |
| $I_{m,f,t}^{D,Im}, I_{m,f,t}^{D,Re}$ | Imaginary and real component of demand current absorbed by bus $m$ at time $t$ |
| $I_{dg,f,t}^{DG,Im}, I_{dg,f,t}^{DG,Re}$ | Imaginary and real component of current injected by diesel generator at period $t$ |
| $I_{gr,f,t}^{GR,Im}, I_{gr,f,t}^{GR,Re}$ | Imaginary and real component of current injected by substation at period $t$ |
| $I_{pv,f,t}^{PV,Im}, I_{pv,f,t}^{PV,Re}$ | Imaginary and real component of current injected by photovoltaic power plant at period $t$ |
| $I_{wt,f,t}^{WT,Im}, I_{wt,f,t}^{WT,Re}$ | Imaginary and real component of current injected by wind turbine at period $t$ |
| $P_{m,t}^{BS}$ | Active power injected by battery system at period $t$ |
| $P_{m,t}^{BS-}$ | Active power discharged from the battery system at period $t$ |
| $P_{m,t}^{BS+}$ | Active power charged to battery system at period $t$ |
| $P_{m,t}^{DG}, P_{m,t}^{PV}, P_{m,t}^{WT}$ | Active power injected by DERs at period $t$ |
| $P_t^{GR}, Q_t^{GR}$ | Active and reactive power injected by substation at period $t$ |
| $\tilde{P}_{m,t}^{PV}, \tilde{P}_{m,t}^{WT}$ | Active power curtailed by photovoltaic plant and wind turbine at period $t$ |
| $\hat{P}_{m,t}^{PV}, \hat{P}_{m,t}^{WT}$ | Available active power at photovoltaic plant and wind turbine in period $t$ |
| $Q_{m,t}^{DG}, Q_{m,t}^{PV}, Q_{m,t}^{WT}, Q_{m,t}^{BS}$ | Reactive power injected by DERs at period $t$ |
| $V_{m,f,t}^{Re}, V_{m,f,t}^{Im}$ | Imaginary and real component of voltage at bus $m$, phase $f$, and time step $t$ |

1. **Introduction**

    The urgent need to transition to less polluting energy sources is encouraging nations around the world to promote renewable energy sources [1]. In recent decades, microgrids (MGs) have proven to be one of the most efficient infrastructures for integrating distributed renewable energy resources (DRERs) into the national energy grid [2]. A MG can be defined as a *"group of interconnected loads and distributed energy resources within clearly defined electrical boundaries that acts as a single controllable entity with respect to the grid. A MG can connect and disconnect from the grid to enable it to operate in both grid-connected or island-mode."* [3].

    The large-scale integration of distributed energy resources (DERs) presents considerable challenges to the operation of MGs and the electrical grid. These challenges include bidirectional power flows, stability concerns, reduced inertia, and the unpredictability of renewable generation [4]. To address these issues, appropriate control techniques must be implemented. When these techniques are properly applied, they can yield additional benefits such as cost savings [5], increased efficiency and reliability [6], peak demand reduction [7], and improved resilience [8].

In MGs, control tasks are commonly divided into three levels: primary level, secondary level and tertiary level [9]. Primary-level tasks focus on maintaining stability in the frequency and voltage levels of the various components of the MG. The secondary level compensates for any deviations in voltage and power levels caused by the primary level. Finally, the tertiary level manages the power flow between the MG and the main grid, ensuring optimal economic operation of the MG through an energy management system (EMS).

The primary task of the energy management system (EMS) is to determine the optimal power injections from dispatchable generators and to manage energy storage systems (ESS) effectively to achieve predefined objectives [10]. There are several approaches to addressing the EMS problem. Regarding the treatment of the electrical grid, one option is to include all electrical constraints to ensure that the solution found is implementable [11]. Another option is to disregard the effects of the electrical grid on the solution [12]. This simplification can be useful in certain cases where the impact of the grid is minimal [13], or when the focus is solely on optimizing local energy resources without considering grid interactions [14]. However, for a more accurate and realistic EMS in MGs the influence of the electrical grid must be considered [15]. Unlike transmission systems, which operate under the assumption of a nearly perfect balance between phases due to their meshed topology and the high-voltage levels involved, MGs exhibit significant imbalances that must be properly modeled and addressed [16]. These imbalances arise primarily from the unbalanced nature of loads, which may include a mixture of single-phase and three-phase connections, as well as from the diverse configurations of electrical lines and distributed energy resources [10]. Moreover, in MGs, factors such as asymmetrical conductor impedances, unbalanced voltage drops, and the integration of single-phase distributed generation units further contribute to the complexity of power flow analysis [9]. As a result, employing a three-phase unbalanced model enables a more precise representation of system behavior, ensuring that voltage profiles and power losses, are accurately captured [15]. This level of detail is crucial for developing effective EMS strategies that optimize energy usage while maintaining system reliability and power quality. On the other hand, the inclusion of unbalanced effects increases the problem's size and complexity. Not surprisingly, most of the research conducted in recent years has employed a single-phase equivalent as the network model [17], leaving the field of three-phase unbalanced MGs largely unexplored. Table 1 summarizes the most prominent works exploring energy management systems in three-phase unbalanced MGs. In the same way, a dynamic lookahead allows the EMS to anticipate future energy demand, generation, and grid conditions, enabling more informed decision-making. It can incorporate real-time forecasting techniques, optimizing energy dispatch strategies while mitigating the risks associated with power fluctuations [13]. Moreover, it facilitates more effective coordination of distributed energy resources (DERs), storage systems, and demand response programs by continuously updating decisions as new information becomes available [18]. Surprisingly, most studies on EMS for microgrids have focused on two-stage programs or static solutions, leaving dynamic lookahead approaches largely underexplored (see Table 1). One of the first studies that attempted to solve the EMS for unbalanced three-phase MGs was conducted by Olivares et al. [19] in 2014. In this seminal work, the EMS problem was decomposed into a unit commitment problem (MILP) and an optimal power flow problem (NLP). The test system used was an isolated 16-bus MG, which included three diesel generators, battery systems, fuel cells, and renewable generation sources. However, this study did not consider the effects of uncertainty, nor was an evaluation conducted on the performance of the EMS in the face of the inherent stochastic variables of the problem.

Three years later, in 2017, Carpinelli et al. [20] proposed a multi-objective optimization program to solve the EMS problem in unbalanced three-phase MGs. The objectives were to reduce voltage imbalance, minimize energy costs, reduce peak demand, decrease electrical losses, and mitigate voltage deviations. A nonlinear deterministic power-based was used, with a test system consisting of a MG located in southern Italy, which included solar panels, data centers equipped with lithium-ion batteries, electric vehicle charging stations, and conventional loads, in a total of 16 buses. A one-day planning window was employed, along with forecasts of loads, renewable generation, and the connection times of electric vehicles. However, no analysis was conducted on the impact of stochasticity on the solution or the effect of the planning window.

In 2018, Hong et al. [21] proposed three strategies for DC MG operation in distribution systems: loss reduction, full unbalance compensation, and partial unbalance compensation using a nonlinear power-based model. The IEEE 123-bus system was used as the test system. However, the study did not incorporate the effect of stochasticity. One year later, Giraldo et al. [10] presented a nonlinear current-based optimization program to solve the energy management problem in unbalanced three-phase MGs. The focus of that work was on modeling synchronous machines under unbalanced operation; renewable energy sources and battery systems were also considered. The objective of the optimization program was to minimize operating costs. The IEEE 123-bus system was used as a test bed. It was demonstrated that disregarding the coupling effects between the lines in a three-phase system could result in overly optimistic solutions; however, the effects of stochasticity in loads, renewable sources, and prices were not considered.

Vergara et al. [22] presented in 2020 a two-stage stochastic mixed-integer nonlinear optimization power-based model for the EMS problem in three-phase islanded MGs. The test system incorporated wind turbines, battery systems, and 25 buses. Stochasticity in power generation and demand was addressed through a scenario generation approach.

Subsequently, in 2021, Castrillon et al. [23] presented a two-stage stochastic mixed-integer second-order cone power-based optimization program to solve the EMS for unbalanced three-phase MGs. The model incorporated dispatchable generators, renewable generators such as wind turbines and photovoltaic systems. The test system used in that study consisted of 25 buses, and stochasticity in both demand and renewable generation was considered. The impact of stochasticity and the decisions obtained were evaluated using Monte Carlo simulations.

The same year, Silva et al. [24] proposed a two-stage stochastic mixed-integer linear power-based optimization program to solve the EMS for unbalanced three-phase MGs. The proposed model considers stochasticity in nodal power demands, renewable generation, and the voltage reference at the point of common coupling. The test system incorporated photovoltaic generation, energy storage systems, electric vehicle chargers, and non-renewable generators. A one-day planning horizon was considered. Even though, a rolling horizon approach was not employed, which would have allowed for decision adjustments as new information became available.

In 2022, Farhad et al. [11] presented a two-stage stochastic current-based nonlinear optimization program to solve the EMS for unbalanced three-phase MGs. The model considered stochasticity in both power generation and demand, as well as energy prices. The test system used was the modified IEEE 34-bus system, incorporating electric vehicle charging stations, wind turbines, photovoltaic systems, diesel generators, and battery systems.

Around the same time, Wang et al. [25] developed a rolling horizon power-based second-order cone optimization program. Stochasticity in loads, renewable generation, and energy prices were considered using a robust optimization approach. The length of the prediction window was set to three hours; however, a study on the influence of the prediction window size on the quality of the obtained solutions was not conducted.

Silva et al. [26] proposed an IoT-based energy management system for the optimal operation of unbalanced three-phase MGs. The proposed architecture incorporated a mixed-integer linear optimization program presented by [24] to determine the day-ahead dispatch of energy sources; stochasticity in loads and renewable generation were addressed through a scenario generation approach. The proposed architecture was tested using the Typhoon HIL platform with real data from a MG at UNICAMP University in São Paulo, which includes a photovoltaic system, a diesel generation unit, and a battery system.

In 2024, Santos et al. [27] presented a methodology for jointly optimizing the sizing and operation of an unbalanced three-phase MG. The proposed EMS minimizes the operating costs of the MG through a two-stage stochastic power-based optimization program based on the work of [24], disregarding the effect of price stochasticity. The sizing of the MG was performed using HOMER Pro software. The proposed strategy was demonstrated to be effective in terms of investment costs for constructing the MG compared to the default dispatch algorithms used by HOMER Pro.

The analysis of the relevant literature on the energy management problem in unbalanced three-phase MGs reveals the prevalence of two-stage stochastic nonlinear or two-stage stochastic mixed integer linear models, as well as power-based models. Concerning two-stage stochastic models, although this methodology has been widely used in the field of stochastic programming, rolling horizon algorithms have the advantage of adjusting decisions based on the latest available information, making them highly responsive to real-time changes and uncertainties [28]. This characteristic is particularly valuable in MG operation, where environmental and load conditions may change rapidly. Another advantage of rolling horizon algorithms is their reduced computational requirements, achieved by breaking down the problem into smaller, more manageable subproblems solved sequentially, rather than addressing a large-scale two-stage optimization program [29].

In the context of power-based optimization models for three-phase unbalanced MGs, previous studies (e.g., [30]) have experimentally demonstrated that the linearization of current-based models generally yields lower error magnitudes compared to that of power-based models. Nevertheless, the linearization approach applied to current-based models in [30] does not facilitate the inclusion of active and reactive power injections as decision variables, which is crucial for accurately modeling the power curtailment strategies commonly employed in modern MG systems.

Moreover, the development of a purely linear model enables the application of various decomposition techniques extensively used in stochastic optimization frameworks. These include classical Bender's decomposition[31], Nested Benders decomposition, and Stochastic Dual Dynamic Programming (SDDP) [32], [14] among others. Such decomposition methods are particularly beneficial for handling the computational complexity associated with large-scale, scenario-based formulations, as they allow for the division of the problem into smaller, more manageable subproblems. This capability not only enhances computational efficiency but also broadens the scope for incorporating uncertainty and variability in power demand and generation within the optimization process.Considering the two key points in the previous paragraph, the aims of this research are formulated as follows:

- This work introduces an advanced rolling horizon algorithm (RL) specifically designed for MG operations. The rolling-horizon optimization framework that dynamically adapts to stochastic renewable generation, demand fluctuations, and energy price volatility, reducing operational costs by 17% compared to conventional two-stage stochastic methods.
- It also introduces two hyperparameters, the prediction window size and the discount factor. These hyperparameters can be optimized by the well stablished methods for hyperparameters optimization. For instance, 25 Optuna [33] trials identify an 11-hour prediction window and discount factor ($\beta=0.997$) as optimal for balancing forecast adaptability and computational efficiency.
- We propose three optimization models derived from the nonlinear current-based model for unbalanced networks. Two of these models are convex, offering well-known convergence properties to the optimal solution. The convex model achieves the highest performance, while linear approximations (4-sided polygon) retain 98% of its accuracy with sub-second solve times, enabling real-time decision-making.

*Table 1. Summary of Research in the Field of Three-Phase Unbalanced MGs*

| | Santos et al., 2024 [27] | Silva et al., 2023 [26] | Wang et al. 2022 [25] | Farhad et al., 2022 [11] | Silva et al., 2021 [24] | Castrillon et al., 2021 [23] | Vergara et al., 2020 [22] | Giraldo et al., 2019 [10] | Hong et al., 2018 [34] | Carpinelli et al., 2017 [20] | Olivares et al., 2014 [19] |
|---|---|---|---|---|---|---|---|---|---|---|---|
| Solution method | D | TSSM | Robust optimization | TSSM | TSSM | TSSM | TSSM | D | D | D | D |
| Formulation | NLP | MILP | SOCP | NLP | MINLP/MILP | MINLP | MINLP/MILP | NLP | Loss reduction | NLP | MINLP/MILP |
| Number of scenarios | 1 | 27 | - | 10 | 27 | 10 | 40 | 1 | 1 | 1 | 1 |
| Three-phase unbalanced power flow | ✓ | ✓ | ✓ | ✓ | ✓ | ✓ | ✓ | ✓ | ✓ | ✓ | ✓ |
| Current based power flow | ✗ | ✗ | ✗ | ✓ | ✗ | ✗ | ✗ | ✓ | ✗ | ✗ | ✓ |
| Stochasticity in loads | ✗ | ✓ | ✓ | ✓ | ✓ | ✓ | ✗ | ✗ | ✗ | ✗ | ✗ |
| Stochasticity in renewable generation | ✗ | ✓ | ✓ | ✓ | ✓ | ✓ | ✗ | ✗ | ✗ | ✗ | ✗ |
| Stochasticity in energy prices | ✗ | ✗ | ✓ | ✓ | ✗ | ✗ | ✗ | ✗ | ✗ | ✗ | ✗ |
| Diesel generators | ✓ | ✓ | ✗ | ✓ | ✓ | ✓ | ✓ | ✓ | ✗ | ✓ | ✓ |
| Energy storage systems | ✓ | ✓ | ✓ | ✓ | ✓ | ✓ | ✓ | ✓ | ✓ | ✓ | ✓ |
| Renewable energy curtailment | ✗ | ✗ | ✗ | ✓ | ✗ | ✓ | ✓ | ✗ | ✗ | ✗ | ✗ |
| Rolling horizon | ✗ | ✗ | ✓ | ✗ | ✗ | ✗ | ✗ | ✗ | ✗ | ✗ | ✗ |

D: Deterministic    TSSM: Two-stage stochastic model

- Significant comparisons are presented to quantify the improvements offered by our approaches: The first comparison assesses the benefits of using a more accurate network model over the traditional balanced single-phase equivalent, highlighting improvements in both accuracy and reliability of the network operation. The second comparison evaluates the advantages of employing a rolling horizon approach to manage uncertainties against the traditional two-stage stochastic model, demonstrating the rolling horizon's enhanced capability in adapting to changing conditions and its impact on operational effectiveness.

The remainder of this paper is structured as follows: section - addresses the problem formulation in terms of state variables, actions, and the transition function, along with the development of various techniques to convexify and linearize it. Section 3 presents the results obtained for the methodologies proposed in the previous section. Section 4 summarizes the main conclusions of this work, and finally, section 5 outlines potential directions for future research.

## 2. Problem formulation

This section presents the general methodology for energy management in a three-phase unbalanced MG, emphasizing its stochastic and dynamic nature. Specifically, Subsection 1.1 describes the essence of the Rolling Horizon (RH) algorithm and motivates its use in MG operation. Subsectionm1.2 introduces the stochastic dynamic framework, including the definition of state vectors, control actions, and random variables. Subsection 1.3 details the mathematical optimization program that embodies the model for decision making (i.e., how the system selects actions). Finally, Subsection 1.4 describes the environment, modeled via a nonlinear power-flow formulation, where actual MG behavior is simulated at each time step when decisions and stochastic realizations occur.

*2.1. Rolling horizon algorithm*

A key contribution of this work is the adoption of a Rolling Horizon (RH) approach to address the MG energy management problem under uncertainty. Let us assume that the MG operates at discrete time steps $t = 0, 1, \ldots, T_{final}$, where $T_{final}$ is the end of the planning horizon (e.g., 24h). At each time step $t'$, the system:

1. Collects the most up-to-date forecasts of uncertain variables---such as renewable generation, energy prices, and load demands---for a prediction window of length $T_W$ (hours) [18].
2. Solves an optimization program (the *model*), which proposes actions (e.g., power injections, storage charges/discharges) for every time step in the interval $T = [t', t' + T_W]$.
3. Implements only the portion of those actions corresponding to the current control step (or possibly a smaller control window) [35].
4. Advances time by one step, i.e., $t' \leftarrow t' + 1$, and *updates* the forecasts and realized stochastic variables. The cycle then repeats.

Figure 1 conceptually illustrates this procedure, showing how the decision process is updated every hour (or every control interval) with fresh forecasts. This scheme is beneficial for two main reasons:

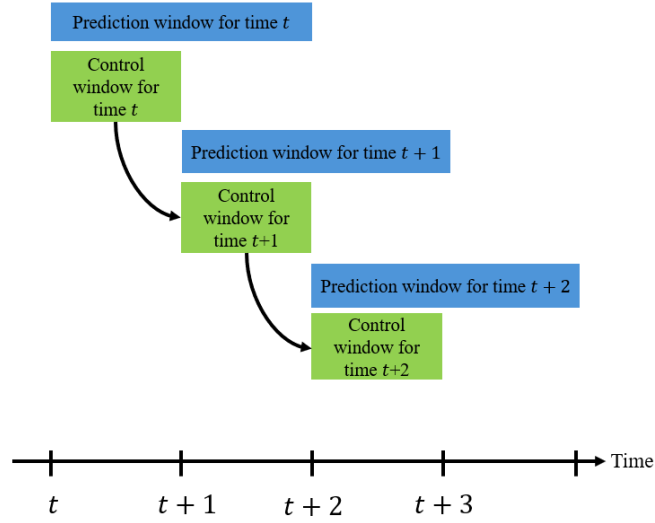

*Figure 1. Illustrative scheme of the Rolling Horizon (RH) algorithm for MG energy management. At each time step $t'$, the optimization program utilizes predictions for a window $[t', t' + T_W]$, but only actions for the immediate control interval are implemented*

- Unlike a single-shot two-stage stochastic model, the RH approach adjusts its decisions as new information arrives, mitigating the risk of basing all decisions on inaccurate or outdated scenarios.
- Instead of solving a large-scale optimization for the entire horizon at once, the problem is *decomposed* into smaller subproblems of manageable size, making it more tractable to solve in near real-time.

*2.2. Stochastic Dynamic Setup: States and Actions*

We cast the MG operation as a sequential (dynamic) and stochastic decision-making problem. Following standard notations in stochastic programming and Markov decision processes [36], for the energy management problem in unbalanced three-phase MGs we define:

**State Vector:**

The state vector is defined in equation (1), which comprises the vector $\overrightarrow{EBS_{t'}}$ representing the state of charge levels for various energy storage systems within the MG; its size is $|BS|$.

$$\overrightarrow{S_{t'}} = \left[ \overrightarrow{EBS_{t'}}, \overrightarrow{PV}_{t':t'+T_W}, \overrightarrow{WV}_{t':t'+T_W}, \overrightarrow{P^D}_{t':t'+T_W}, \overrightarrow{Q^D}_{t':t'+T_W}, \overrightarrow{C^{GR}}_{t':t'+T_W} \right] \quad (1)$$

The vector $\overrightarrow{PV}_{t':t'+T_W}$ represents the predicted solar energy generation for each bus bar of the MG where a solar farm is installed, for the prediction window $[t', t' + T_W]$; its size is $|PV| \cdot T_W$. The vector $\overrightarrow{WV}_{t':t'+T_W}$ represents the predicted wind speed at each bus bar of the

MG where wind turbines are installed, for the prediction window $[t', t' + T_W]$; its size is $|WT| \cdot T_W$. The vector $\overrightarrow{P^D}_{t':t'+T_W}$ represents the active power demand predictions at each bus bar of the MG, for the prediction window $[t', t' + T_W]$; its size is $|\Omega| \cdot T_W$. Similarly, the vector $\overrightarrow{Q^D}_{t':t'+T_W}$ represents the reactive power demand predictions at each bus bar of the MG, for the prediction window $[t', t' + T_W]$, and its size is also $|\Omega| \cdot T_W$. Finally, the vector $\overrightarrow{C^{GR}}_{t':t'+T_W}$ represents the energy prices' predictions for the prediction window $[t', t' + T_W]$, and its size is $T_W$.

**Action Vector:**

The action vector for time step $t'$ is defined in equation ( 2); these actions correspond to the operational decisions that the MG operator must set for the immediate time step $t' + 1$, including the following: the vector $\overrightarrow{PBS_{t'}}$ represents the power charged and discharged, for the control window, for each energy storage system presented in the MG; its size is $|BS|$.

$$\overrightarrow{x_{t'}} = \left[\overrightarrow{PBS_{t'}}, \overrightarrow{P_{t'}^{DG}}, \overrightarrow{Q_{t'}^{DG}}, \overrightarrow{\widetilde{P}_{t'}^{PV}}, \overrightarrow{\widetilde{P}_{t'}^{WT}}, \overrightarrow{Q_{t'}^{PV}}, \overrightarrow{Q_{t'}^{WT}}, \overrightarrow{Q_{t'}^{BS}}\right] \qquad (2)$$

In the same way, the vector $\overrightarrow{P_{t'}^{DG}}$ and $\overrightarrow{Q_{t'}^{DG}}$ represents the active and reactive power generation, for the control window, for each diesel generator unit present in the MG; its size is $|DG|$. The vectors $\overrightarrow{\widetilde{P}_{t'}^{PV}}$ and $\overrightarrow{\widetilde{P}_{t'}^{WT}}$ represents the power curtailments, for control window, for solar farms and wind turbines present in the MG; their sizes are $|PV|$ and $|WT|$ respectively. Likewise, $\overrightarrow{Q_{t'}^{PV}}$ and $\overrightarrow{Q_{t'}^{WT}}$ are the reactive power generations, for the control window, for solar farms and wind turbines present in the MG; their sizes are also $|PV|$ and $|WT|$. Finally, $\overrightarrow{Q_{t'}^{BS}}$ is the reactive power generation, for the control window, for the energy storage systems; its size is $|BS|$.

**Transition Function:**

Once an action vector $\overrightarrow{x_{t'}}$ is applied at time $t'$ and the random variables (e.g., realized PV/wind generation, prices, and demands) are revealed as $\overrightarrow{W_{t'}}$, the MG state evolves from $\overrightarrow{S_{t'}}$ to a new state $\overrightarrow{S_{t'+1}}$. Formally, this relation is established through equation (3)

$$\overrightarrow{S_{t'+1}} = \mathrm{E}(\overrightarrow{S_{t'}}, \overrightarrow{x_{t'}}, \overrightarrow{W_{t'}}) \qquad (3)$$

In this work, $\mathrm{E}(\cdot)$ is represented by a power flow analysis capturing the three-phase unbalanced nature of the network [11], as explained in Subsection 2.4.

*2.3. Model based on mathematical programming*
A model is a solution for the problem stated in the previous sections [18]. In this sense, the actions can be expressed as

$$\overrightarrow{x_{t'}} = \Pi(\overrightarrow{S_{t'}})$$

In this work, model Π is an optimization program, based on the work of [11]; at each time $t$, the model Π is given by a mathematical program that seeks to minimize the expected operational cost of the MG over the window $T = [t', t' + T_W]$. For simplicity of presentation, one can view it as a deterministic problem using the available forecasts at time $t'$ (possibly with scenario-based or point forecasts). The core model uses the following objective and constraints based on the two-stage stochastic program proposed by [11], with some modifications that will be discussed in brief:

- *Objective function*

The objective function (4), aims to minimize the operating cost of the MG. The first term of the objective function, $\sum_{T=0}^{T=T_W} \beta^T \cdot \Delta t \cdot C_t^{Gr} \cdot P_t^{Gr}$ corresponds to the cost of energy purchased from the main grid, while the second term $\sum_{dg \in DG} \sum_{T=0}^{T=T_W} \beta^T \cdot \Delta t \cdot C_t^{dg} \cdot P_t^{dg}$ represents the operating cost of the diesel generators. The discount factor $\beta$ is introduced by us to mitigate the effect of optimistic decisions obtained when using rolling horizon algorithms, as described in [18].

$$\min \sum_{t \in T} \beta^i \cdot \Delta t \cdot C_t^{Gr} \cdot P_t^{Gr} + \sum_{dg \in DG} \sum_{t \in T} \beta^i \cdot \Delta t \cdot C_t^{dg} \cdot P_t^{dg} \tag{4}$$

- *Constraints*

The constraints are the current based representation of the three-phase unbalanced power flow for an electrical network.

- *Voltage drops and Kirchhoff law constraints*

Constraints (5) and (6) relate de voltage drop across a transmission line to the electrical current flowing through it, as extensively explained in [30]; $R_{mn,f,h}$ represents the resistance between buses $m$ and $n$ in phase $h$, while $X_{mn,h}$ denotes the corresponding reactance. The well-known Kirchhoff's current law is expressed in constraints (7) and (8), ensuring that the total sum of currents entering a bus equal zero. Constraints (9) and (10) impose technical limits to nodal voltages and currents in the MG.

$$V_{m,f,t}^{Re} - V_{n,f,t}^{Re} = \sum_{h \in F} \left( R_{mn,h} \cdot I_{mn,h,t}^{Re} - X_{mn,h} \cdot I_{mn,h,t}^{Im} \right) \forall m \in N, f \in H, t \in T \tag{5}$$

$$V_{m,f,t}^{Im} - V_{n,f,t}^{Im} = \sum_{h \in F} \left( X_{mn,h} \cdot I_{mn,h,t}^{Re} + R_{mn,h} \cdot I_{mn,h,t}^{Im} \right) \forall m \in N, f \in H, t \in T \tag{6}$$

$$\sum_{gr \in GR|gr=m} I_{gr,f,t}^{GR,Re} + \sum_{pv \in PV|pv=m} I_{pv,f,t}^{PV,Re} + \sum_{wt \in WT|wt=m} I_{wt,f,t}^{WT,Re} + \sum_{dg \in DG|dg=m} I_{dg,f,t}^{DG,Re} - \sum_{bs \in BS|bs=m} I_{bs,f,t}^{BS,Re} \tag{7}$$
$$+ \sum_{km \in L} I_{km,f,t}^{Re} - \sum_{mn \in L} I_{mn,f,t}^{Re} - \left( \sum_{km \in L} B_{km,f} + \sum_{mn \in L} B_{mn,f} \right) \cdot \frac{V_{m,f,t}^{Im}}{2} = I_{m,f,t}^{D,Re} \quad \forall m \in N, f \in H, t \in T$$

$$\sum_{gr \in GR|gr=m} I_{gr,f,t}^{GR,Im} + \sum_{pv \in PV|pv=m} I_{pv,f,t}^{PV,Im} + \sum_{wt \in WT|wt=m} I_{wt,f,t}^{WT,Im} + \sum_{dg \in DG|dg=m} I_{dg,f,t}^{DG,Im} - \sum_{bs \in BS|bs=m} I_{bs,f,t}^{BS,Im} \tag{8}$$
$$+ \sum_{km \in L} I_{km,f,t}^{Im} - \sum_{mn \in L} I_{mn,f,t}^{Im} - \left( \sum_{km \in L} B_{km,f} + \sum_{mn \in L} B_{mn,f} \right) \cdot \frac{V_{m,f,t}^{Re}}{2} = I_{m,f,t}^{D,Im} \quad \forall m \in N, f \in H, t \in T$$

$$\underline{V}^2 \leq {V_{m,f,t}^{Re}}^2 + {V_{m,f,t}^{Im}}^2 \leq \overline{V}^2 \quad \forall\, m \in N, f \in H, t \in T \tag{9}$$

$$ {I_{mn,f,t}^{Re}}^2 + {I_{mn,f,t}^{Im}}^2 \leq \overline{I_{mn}}^2 \quad \forall\, mn \in L, f \in H, t \in T \tag{10}$$

- *Constraints related to electrical loads*

Constraints (11) and (12) relate de active and reactive power demand to real and imaginary nodal voltages and real and imaginary demand currents, as suggested in [30].

$$P_{m,f,t}^{D} = V_{m,f,t}^{Re} \cdot I_{m,f,t}^{D,Re} + V_{m,f,t}^{Im} \cdot I_{m,f,t}^{D,Im} \quad \forall\, m \in N, f \in H, t \in T \tag{11}$$

$$Q_{m,f,t}^{D} = -V_{m,f,t}^{Re} \cdot I_{m,f,t}^{D,Im} + V_{m,f,t}^{Im} \cdot I_{m,f,t}^{D,Re} \quad \forall\, m \in N, f \in H, t \in T \tag{12}$$

- *Constraints related to PVs, WT and DGs*

Similarly, constraints (13) and (14) relate the active and reactive power generation of photovoltaic panels with real and imaginary nodal voltages and the corresponding injected currents. The apparent power limit of the photovoltaic panels installed in the MG is enforced by constraint (15), while constraint (16) guarantees a specified power factor value for the photovoltaic panels. The active power curtailment method, commonly used to limit the power generation of renewable sources [11], is modeled trough constraint (17); in this way, the total active power available, $\hat{P}_{pv,t}^{PV}$, can be divided among the actual power injected into the MG, $P_{pv,t}^{PV}$ and the amount of power curtailment $\tilde{P}_{pv,t}^{PV}$. Finally, constraint (18) imposes limits on the reactive power generation for the photovoltaic panels.

In a similar manner, constraints (19) - (24) model the behavior of the wind turbines present in the MG. The operation of the diesel generators is simulated by constraints (25) - (29), which exhibit the same pattern that the constraints used for photovoltaic panels and wind turbines.

$$\frac{P_{m,t}^{PV}}{3} = V_{m,f,t}^{Re} \cdot I_{m,f,t}^{PV,Re} + V_{m,f,t}^{Im} \cdot I_{m,f,t}^{PV,Im} \quad \forall\, m \in PV, f \in H, t \in T \tag{13}$$

$$\frac{Q_{m,t}^{PV}}{3} = -V_{m,f,t}^{Re} \cdot I_{m,f,t}^{PV,Im} + V_{m,f,t}^{Im} \cdot I_{m,f,t}^{PV,Re} \quad \forall\, m \in PV, f \in H, t \in T \tag{14}$$

$${P_{pv,t}^{PV}}^2 + {Q_{pv,t}^{PV}}^2 \leq \overline{S_{pv}^{PV}}^2 \quad \forall\, pv \in PV, t \in T \tag{15}$$

$$Q_{pv,t}^{PV} \leq P_{pv,t}^{PV} \cdot \tan\!\left(\operatorname{acos}\!\left(pf_{pv}^{PV}\right)\right) \quad \forall\, pv \in PV, t \in T \tag{16}$$

$$\hat{P}_{pv,t}^{PV} = P_{pv,t}^{PV} + \tilde{P}_{pv,t}^{PV} \quad \forall\, pv \in PV, t \in T \tag{17}$$

$$\underline{Q_{pv}^{PV}} \leq Q_{pv,t}^{PV} \leq \overline{Q_{pv}^{PV}} \quad \forall \, pv \in PV, t \in T \tag{18}$$

$$\frac{P_{m,t}^{WT}}{3} = V_{m,f,t}^{Re} \cdot I_{m,f,t}^{WT,Re} + V_{m,f,t}^{Im} \cdot I_{m,f,t}^{WT,Im} \quad \forall \, m \in WT, f \in H, t \in T \tag{19}$$

$$\frac{Q_{m,t}^{WT}}{3} = -V_{m,f,t}^{Re} \cdot I_{m,f,t}^{WT,Im} + V_{m,f,t}^{Im} \cdot I_{m,f,t}^{WT,Re} \quad \forall \, m \in WT, f \in H, t \in T \tag{20}$$

$$P_{wt,t}^{WT\,2} + Q_{wt,t}^{WT\,2} \leq \overline{S_{wt}^{WT}}^2 \quad \forall \, wt \in WT, t \in T \tag{21}$$

$$Q_{wt,t}^{WT} \leq P_{wt,t}^{WT} \cdot \tan(\text{acos}(pf_{wt}^{WT})) \quad \forall \, wt \in WT, t \in T \tag{22}$$

$$\hat{P}_{wt,t}^{WT} = P_{wt,t}^{WT} + \tilde{P}_{wt,t}^{WT} \quad \forall \, wt \in WT, t \in T \tag{23}$$

$$\underline{Q_{wt}^{WT}} \leq Q_{wt,t}^{WT} \leq \overline{Q_{wt}^{WT}} \quad \forall \, wt \in WT, t \in T \tag{24}$$

$$\frac{P_{m,t}^{DG}}{3} = V_{m,f,t}^{Re} \cdot I_{m,f,t}^{DG,Re} + V_{m,f,t}^{Im} \cdot I_{m,f,t}^{DG,Im} \quad \forall \, m \in DG, f \in H, t \in T \tag{25}$$

$$\frac{Q_{m,t}^{DG}}{3} = -V_{m,f,t}^{Re} \cdot I_{m,f,t}^{DG,Im} + V_{m,f,t}^{Im} \cdot I_{m,f,t}^{DG,Re} \quad \forall \, m \in DG, f \in H, t \in T \tag{26}$$

$$P_{dg,t}^{DG\,2} + Q_{dg,t}^{DG\,2} \leq \overline{S_{dg}^{DG}}^2 \quad \forall \, dg \in DG, t \in T \tag{27}$$

$$Q_{dg,t}^{DG} \leq P_{dg,t}^{DG} \cdot \tan(\text{acos}(pf_{dg}^{DG})) \quad \forall \, dg \in DG, t \in T \tag{28}$$

$$\underline{Q_{dg}^{DG}} \leq Q_{dg,t}^{DG} \leq \overline{Q_{dg}^{DG}} \quad \forall \, dg \in DG, t \in T \tag{29}$$

- *Constraints related to battery system*

Constraints (30) - (37) represent the behavior of battery systems installed in the MG; constraints (30) and (31) relate the active and reactive power injections from the battery to real and imaginary nodal voltages and current injections. Constraint (32) defines two continuous positive variables for charging ($P_{m,t}^{BS+}$) and discharging ($P_{m,t}^{BS-}$) active power from the battery, and constraint (33) relates the energy stored at the battery at time $t$ with the energy stored at the battery at $t-1$ and power charge and discharged, taking into account charging and discharging efficiencies [11]. Constraint (34) has been introduced in this work to limit the reactive power generation by the specification of a minimum power factor for the battery system. Finally, constraints (35) - (37) represent the operational limits of the battery system.

$$\frac{P_{m,t}^{BS}}{3} = V_{m,f,t}^{Re} \cdot I_{m,f,t}^{BS,Re} + V_{m,f,t}^{Im} \cdot I_{m,f,t}^{BS,Im} \quad \forall\, m \in BS, f \in H, t \in T \tag{30}$$

$$\frac{Q_{m,t}^{BS}}{3} = -V_{m,f,t}^{Re} \cdot I_{m,f,t}^{BS,Im} + V_{m,f,t}^{Im} \cdot I_{m,f,t}^{BS,Re} \quad \forall\, m \in BS, f \in H, t \in T \tag{31}$$

$$P_{m,t}^{BS} = P_{m,t}^{BS+} - P_{m,t}^{BS-} \quad \forall\, m \in BS, t \in T \tag{32}$$

$$E_{m,t}^{BS} = E_{m,t-1}^{BS} + \Delta t \left( P_{m,t}^{BS+} \cdot \eta_{bs}^{BS} - \frac{P_{m,t}^{BS-}}{\eta_{bs}^{BS}} - E_{m,t}^{BS} \cdot \beta_{bs}^{BS} \right) \quad \forall\, m \in BS,\, t \in T \tag{33}$$

$$|Q_{m,t}^{BS}| \leq |P_{m,t}^{BS} \cdot \tan(\operatorname{acos}(pf_{bs}^{BS}))| \quad \forall\, m \in BS, t \in T \tag{34}$$

$$\underline{E}_m^{BS} \leq E_{m,t}^{BS} \leq \overline{E}_m^{BS} \quad \forall\ m \in BS, t \in T \tag{35}$$

$$0 \leq P_{m,t}^{BS+} \leq \overline{P}_m^{BS+} \quad \forall\ m \in BS,\, t \in T \tag{36}$$

$$0 \leq P_{m,t}^{BS-} \leq \overline{P}_m^{BS-} \quad \forall\ m \in BS,\, t \in T \tag{37}$$

- *Constraints related to substation*

Constraints (38) and (39) define the relationship between active and reactive power from the main grid, the voltage level and the injected current in the substation. The amount of power that can be interchanged (both sold and purchased) with the main grid is constrained by the capacity of the transformer, as expressed in constraint (40).

$$P_{t'}^{Gr} = \sum_{f \in H} (V_{gr,f,t'}^{Re} \cdot I_{gr,f,t'}^{GR\ Re} + V_{gr,f,t'}^{Im} \cdot I_{gr,f,t'}^{GR\ Im}) \quad \forall\, t \in T \tag{38}$$

$$Q_{t'}^{Gr} = \sum_{f \in H} (-V_{gr,f,t'}^{Re} \cdot I_{gr,f,t'}^{GR\ Im} + V_{gr,f,t'}^{Im} \cdot I_{gr,f,t'}^{GR\ Re}) \quad \forall\, t \in T \tag{39}$$

$$P_{t'}^{Gr^2} + Q_{t'}^{Gr^2} \leq S_{Tr}^2 \quad \forall\, t \in T \tag{40}$$

### 2.3.1. Nonconvex quadratic model

Because of the bilinear terms in (11)–(14), (19)–(20), (25)–(26), and (30)–(31), and the squared terms in (9) and (10), the above formulation is nonlinear and nonconvex. Even for small microgrids, the computational resources required to solve this optimization program with off-the-shelf software are often prohibitively high, making it impractical for real-time applications. To make the model tractable, quadratic and linear optimization models are derived below from the original nonlinear optimization program.

By linearizing constraints (11) - (14), (19) - (20), (25) - (26) and (30) - (31) the nonlinear optimization program presented in the previous section can be transformed into a non-convex quadratic optimization problem. Concerning constraints (11) - (14), as exposed in [30], the linearization is achieved by truncating the Taylor expansion for real and imaginary currents in the second term:

$$I_{m,f,t}^{D,Re} = g^* + \frac{\partial g}{\partial V_{m,f,t}^{Re}}^* \left(V_{m,f,t}^{Re} - V_{m,f,t}^{Re\,*}\right) + \frac{\partial g}{\partial V_{m,f,t}^{Im}}^* \left(V_{m,f,t}^{Im} - V_{m,f,t}^{Im\,*}\right) \forall\, m \in N, f \in H, t \in \{0, T_W\} \quad (41)$$

$$I_{m,f,t}^{D,Im} = h^* + \frac{\partial h}{\partial V_{m,f,t}^{Re}}^* \left(V_{m,f,t}^{Re} - V_{m,f,t}^{Re\,*}\right) + \frac{\partial h}{\partial V_{m,f,t}^{Im}}^* \left(V_{m,f,t}^{Im} - V_{m,f,t}^{Im\,*}\right) \forall\, m \in N, f \in H, t \in \{0, T_W\} \quad (42)$$

Functions $g$, $h$, $\frac{\partial g}{\partial V^{Re}}$, $\frac{\partial g}{\partial V^{Im}}$, $\frac{\partial h}{\partial V^{Re}}$ and $\frac{\partial h}{\partial V^{Im}}$ are presented in Annex A. Regarding constraints (19) - (20), (25) - (26) and (30) - (31), since active and reactive powers are decision variables, the methodology proposed by [30] is extended in this work by considering the real and imaginary currents as functions of four variables ($V^{Re}$, $V^{Im}$, $P$ and $Q$). Consequently, the linear expressions consist of four terms, as shown in constraints (43) - (50).

$$I_{m,f,t}^{PV,Re} = g^* + \frac{\partial g}{\partial V_{m,f,t}^{Re}}^* \left(V_{m,f,t}^{Re} - V_{m,f,t}^{Re\,*}\right) + \frac{\partial g}{\partial V_{m,f,t}^{Im}}^* \left(V_{m,f,t}^{Im} - V_{m,f,t}^{Im\,*}\right) + \frac{\frac{\partial g}{\partial P_{m,f,t}^{PV}}^* \left(P_{m,f,t}^{PV} - P_{m,f,t}^{PV\,*}\right)}{3} + \frac{\frac{\partial g}{\partial Q_{m,f,t}^{PV}}^* \left(Q_{m,f,t}^{PV} - Q_{m,f,t}^{PV\,*}\right)}{3} \quad \forall\, m \in PV, f \in H, t \in \{0, T_W\} \quad (43)$$

$$I_{m,f,t}^{PV,Im} = h^* + \frac{\partial h}{\partial V_{m,f,t}^{Re}}^* \left(V_{m,f,t}^{Re} - V_{m,f,t}^{Re\,*}\right) + \frac{\partial h}{\partial V_{m,f,t}^{Im}}^* \left(V_{m,f,t}^{Im} - V_{m,f,t}^{Im\,*}\right) + \frac{\frac{\partial h}{\partial P_{m,f,t}^{PV}}^* \left(P_{m,f,t}^{PV} - P_{m,f,t}^{PV\,*}\right)}{3} + \frac{\frac{\partial h}{\partial Q_{m,f,t}^{PV}}^* \left(Q_{m,f,t}^{PV} - Q_{m,f,t}^{PV\,*}\right)}{3} \quad \forall\, m \in PV, f \in H, t \in \{0, T_W\} \quad (44)$$

$$I_{m,f,t}^{WT,Re} = g^* + \frac{\partial g}{\partial V_{m,f,t}^{Re}}^* \left(V_{m,f,t}^{Re} - V_{m,f,t}^{Re\,*}\right) + \frac{\partial g}{\partial V_{m,f,t}^{Im}}^* \left(V_{m,f,t}^{Im} - V_{m,f,t}^{Im\,*}\right) + \frac{\frac{\partial g}{\partial P_{m,f,t}^{WT}}^* \left(P_{m,f,t}^{WT} - P_{m,f,t}^{WT\,*}\right)}{3} + \frac{\frac{\partial g}{\partial Q_{m,f,t}^{WT}}^* \left(Q_{m,f,t}^{WT} - Q_{m,f,t}^{WT\,*}\right)}{3} \quad \forall\, m \in WT, f \in H, t \in \{0, T_W\} \quad (45)$$

$$I_{m,f,t}^{WT,Im} = h^* + \frac{\partial h}{\partial V_{m,f,t}^{Re}}^* \left(V_{m,f,t}^{Re} - V_{m,f,t}^{Re\,*}\right) + \frac{\partial h}{\partial V_{m,f,t}^{Im}}^* \left(V_{m,f,t}^{Im} - V_{m,f,t}^{Im\,*}\right) + \frac{\frac{\partial h}{\partial P_{m,f,t}^{WT}}^* \left(P_{m,f,t}^{WT} - P_{m,f,t}^{WT\,*}\right)}{3} + \frac{\frac{\partial h}{\partial Q_{m,f,t}^{WT}}^* \left(Q_{m,f,t}^{WT} - Q_{m,f,t}^{WT\,*}\right)}{3} \quad \forall m \in WT, f \in H, t \in \{0, T_W\} \tag{46}$$

$$I_{m,f,t}^{DG,Re} = g^* + \frac{\partial g}{\partial V_{m,f,t}^{Re}}^* \left(V_{m,f,t}^{Re} - V_{m,f,t}^{Re\,*}\right) + \frac{\partial g}{\partial V_{m,f,t}^{Im}}^* \left(V_{m,f,t}^{Im} - V_{m,f,t}^{Im\,*}\right) + \frac{\frac{\partial g}{\partial P_{m,f,t}^{DG}}^* \left(P_{m,f,t}^{DG} - P_{m,f,t}^{DG\,*}\right)}{3} + \frac{\frac{\partial g}{\partial Q_{m,f,t}^{DG}}^* \left(Q_{m,f,t}^{DG} - Q_{m,f,t}^{DG\,*}\right)}{3} \quad \forall m \in DG, f \in H, t \in \{0, T_W\} \tag{47}$$

$$I_{m,f,t}^{DG,Im} = h^* + \frac{\partial h}{\partial V_{m,f,t}^{Re}}^* \left(V_{m,f,t}^{Re} - V_{m,f,t}^{Re\,*}\right) + \frac{\partial h}{\partial V_{m,f,t}^{Im}}^* \left(V_{m,f,t}^{Im} - V_{m,f,t}^{Im\,*}\right) + \frac{\frac{\partial h}{\partial P_{m,f,t}^{DG}}^* \left(P_{m,f,t}^{DG} - P_{m,f,t}^{DG\,*}\right)}{3} + \frac{\frac{\partial h}{\partial Q_{m,f,t}^{DG}}^* \left(Q_{m,f,t}^{DG} - Q_{m,f,t}^{DG\,*}\right)}{3} \quad \forall m \in DG, f \in H, t \in \{0, T_W\} \tag{48}$$

$$I_{m,f,t}^{BS,Re} = g^* + \frac{\partial g}{\partial V_{m,f,t}^{Re}}^* \left(V_{m,f,t}^{Re} - V_{m,f,t}^{Re\,*}\right) + \frac{\partial g}{\partial V_{m,f,t}^{Im}}^* \left(V_{m,f,t}^{Im} - V_{m,f,t}^{Im\,*}\right) + \frac{\frac{\partial g}{\partial P_{m,f,t}^{DG}}^* \left(P_{m,f,t}^{BS} - P_{m,f,t}^{BS\,*}\right)}{3} + \frac{\frac{\partial g}{\partial Q_{m,f,t}^{BS}}^* \left(Q_{m,f,t}^{BS} - Q_{m,f,t}^{BS\,*}\right)}{3} \quad \forall m \in BS, f \in H, t \in \{0, T_W\} \tag{49}$$

$$I_{m,f,t}^{BS,Im} = h^* + \frac{\partial h}{\partial V_{m,f,t}^{Re}}^* \left(V_{m,f,t}^{Re} - V_{m,f,t}^{Re\,*}\right) + \frac{\partial h}{\partial V_{m,f,t}^{Im}}^* \left(V_{m,f,t}^{Im} - V_{m,f,t}^{Im\,*}\right) + \frac{\frac{\partial h}{\partial P_{m,f,t}^{BS}}^* \left(P_{m,f,t}^{BS} - P_{m,f,t}^{BS\,*}\right)}{3} + \frac{\frac{\partial h}{\partial Q_{m,f,t}^{BS}}^* \left(Q_{m,f,t}^{BS} - Q_{m,f,t}^{BS\,*}\right)}{3} \quad \forall m \in BS, f \in H, t \in \{0, T_W\} \tag{50}$$

As a result, by replacing constraints (11) - (14), (19) - (20), (25) - (26) and (30) - (31) by constraints (43) - (50), the original nonlinear nonconvex optimization problem becomes a quadratically constrained optimization problem QCP. Due to the remaining squared constraints (*9*) which are nonconvex, the resulting problem is a nonconvex QCP, more tractable than the original but still posing global optimality challenges for even for small instances.

### 2.3.2. *QCP convex model*

The principal source of nonconvexity in the nonconvex QCP program from section 2.3.1 stems from constraint (9), as illustrated in Figure 2. The constraint $\underline{V}^2 \leq {V_{m,f,t}^{Re}}^2 + {V_{m,f,t}^{Im}}^2$ renders the feasible space nonconvex. To derive a convex optimization program, this constraint can be replaced with a linear constraint, as presented in [37] and [38] and is depicted in Figure 3.

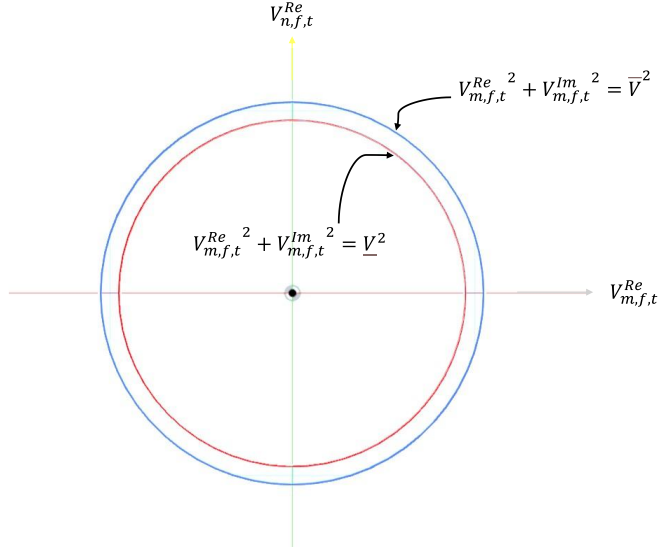

Figure 2. Nonconvex feasible space for voltage constraint

Thus, constraint (9) is replaced by constraints (51) y (52) [16]. In constraint (52), $\theta_f$ is the operation angle for phase $f$ and $\phi$ is the maximum deviation allowed around the operating angle $\theta_f$ [16].

$$V_{m,f,t}^{Re}{}^2 + V_{m,f,t}^{Im}{}^2 \leq \overline{V}^2 \quad \forall\, m \in N, f \in H, t \in T \tag{51}$$

$$V_{m,f,t}^{Im} \leq \frac{\sin(\theta_f + \phi) - \sin(\theta_f - \phi)}{\cos(\theta_f + \phi) - \cos(\theta_f - \phi)} \cdot [V_{m,f,t}^{Re} - \underline{V}\cos(\theta_f - \phi)] + \underline{V}\sin(\theta_f - \phi) \quad \forall\, m \in N, f \in H, t \in T \tag{52}$$

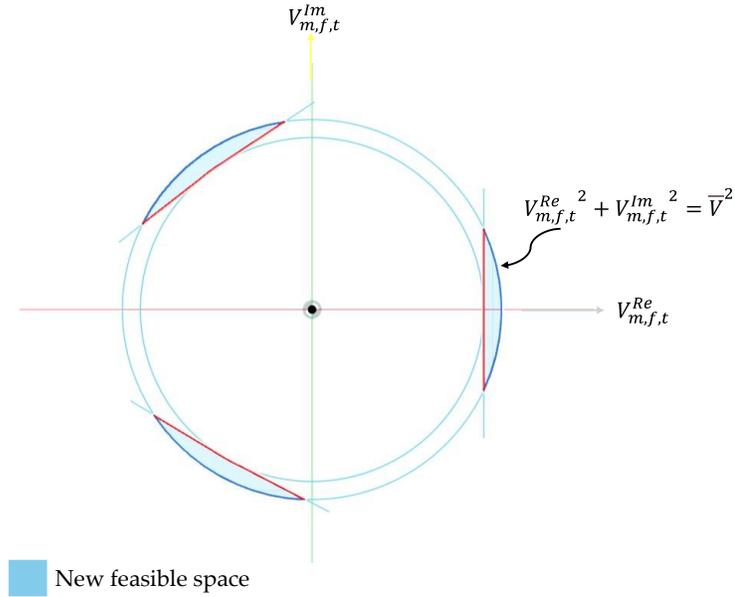

New feasible space

Figure 3. Convex feasible space for voltage constraints

### 2.3.3. Linear model

By linearizing all the quadratic constraints of the general form $X_1^2 + X_2^2 \leq C^2$ (i.e., constraints (51), (10), (15), (21) and (27)) a linear optimization program can be derived. The linearization process involves constructing linear constraints for the edges of a polygon inscribed within the circle defined by $|X_1|^2 + |X_2|^2 \leq C^2$ [37], specifically for the first quadrant [39]. As illustrated in Figure 4, the general quadratic constraint $|X_1|^2 + |X_2|^2 \leq C^2$ is approximated by inscribing a polygon, enabling arbitrary precision at the cost of increasing the number of constraints.

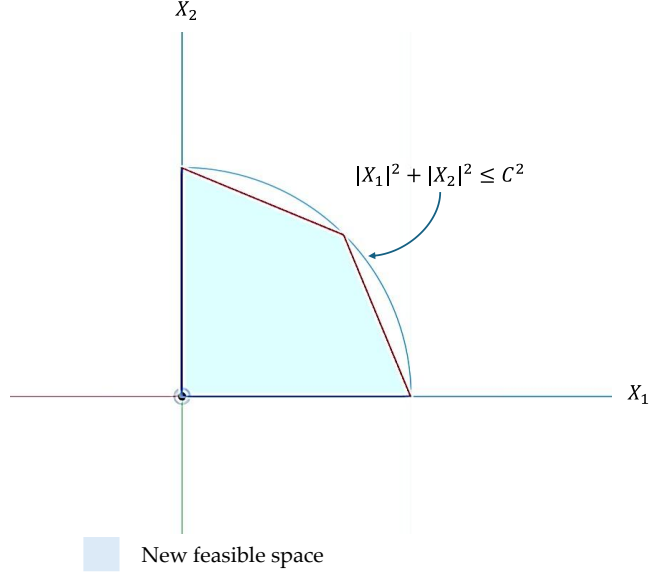

Figure 4. Approximation of the quadratic constraint $X_1^2 + X_2^2 \leq C^2$ by inscribing an 8-faces polygon.

In the special case of constraint (51), which defines the upper limit for the voltage level, it is replaced by a set of $2l$ line segments. Among these $l$ segments are constructed counterclockwise and $l$ segments clockwise, starting from operating angle $\theta_f$, as outlined in constraint (53). The angle $\varphi$ represents the arc angle corresponding to each constructed line segment.

$$V_{m,f,t}^{Im} \leq \frac{\sin(\theta_f + l_i * \varphi) - \sin(\theta_f + (l_i - 1)\varphi)}{\cos(\theta_f + l_i * \varphi) - \cos(\theta_f + (l_i - 1)\varphi)} \cdot [V_{m,f,t}^{Re} - \underline{V}\cos(\theta_f + (l_i - 1)\varphi)] \\ + \underline{V}\sin(\theta_f + (l_i - 1)\varphi) \quad \forall\, m \in N, f \in H, t \in T, i \in [-l, l]$$
(53)

### 2.4. Environment: Nonlinear Power Flow

In this work, the environment is a function E that relates the state of each time stage, and the corresponding actions with the stage of the next time stage:

$$\overrightarrow{S_{t+1}} = \mathrm{E}(\overrightarrow{S_t}, \overrightarrow{x_t}, \overrightarrow{W_t})$$

where $\overrightarrow{W_t}$ is the true realization of the stochastic variables considered in the problem (i.e solar power generation for each bus bar, wind velocity for each bus bar, active power demand for each bus bar, reactive power demand for each bus bar and energy price). In this work, the environment (i.e function E) is based on the power flow analysis proposed by [30]. The behavior of the MG when the set of actions $\vec{x_t}$ and the realizations $\vec{W_t}$ are applied on the environment can be modeled by a free optimization program defined by equations (5) - (14), (19) - (20), (25) - (26) and (30) -(33), plugging in the decisions and the true realizations of the stochastic variables $(\vec{x_t}, \vec{W_t})$. Consequently, the optimization problem is formulated as a nonconvex nonlinear optimization program which can be solved using off-the-shelf software within acceptable time frames without requiring further simplifications, thus enabling the testing of the models outlined in section 2.3 within a nonlinear environment as shown in Figure 5. **Remark**: Implementing E(·)in practice can be done by standard power-flow solvers (e.g., Newton-Raphson for unbalanced networks) or by a small nonlinear optimization problem as in (5) - (14), (19) - (20), (25) - (26) and (30) -(33). This flexible "plug-in" approach allows more accurate physics-based modeling of MGs under uncertain conditions.

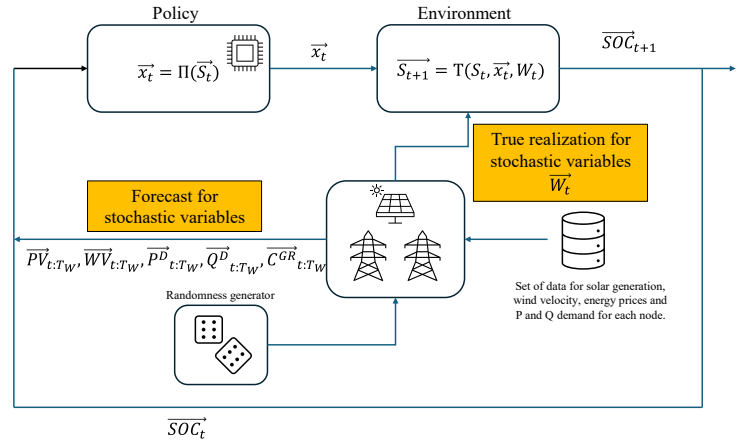

*Figure 5. General scheme of methodology used in this work to simulate the behavior of the MG.*

## 2.5. Hyperparameters in the Rolling Horizon Framework

The proposed rolling horizon algorithm relies on two key hyperparameters that critically influence the performance and responsiveness of the MG energy management:

**Prediction Window Size, $T_W$:** This parameter specifies how many future times steps the optimization program will consider at each control interval. A larger $T_W$ can provide a more comprehensive outlook on future uncertainties and potential system states, yet it also increases forecast errors in distant time steps and the computational burden of each solve. A smaller $T_W$ on the other hand, reduces problem complexity but may result in short-sighted decisions.

**Discount Factor, $\beta$:** This parameter controls how strongly future costs (or rewards) are weighted relative to immediate ones. A discount factor closes to one gives nearly equal importance on all time steps, whereas smaller values prioritize short-term costs over long-term outcomes. Additionally, the discount factor plays a subtle role in preventing so-called rollout bias or overly optimistic future cost estimates, especially under uncertain or partially inaccurate forecasts.

Earlier works in rolling horizon control for microgrids often choose $T_W$ a priori. However, the optimal $T_W$ is generally application-specific and can be tuned using either parametric sweeps or formal hyperparameter search methods (e.g., grid search, Bayesian optimization). In our study, we systematically optimize $T_W$ (alongside with $\beta$) using an open-source library, as described in Section 3.2.

## 3. Results

This section presents the results obtained for the models proposed in Section - aimed at achieving energy management in a three-phase unbalanced MG, with stochasticity in both active and reactive power demand, renewable energy generation, and energy prices. Subsection 3.1 describes the test system used in this study and provides details on the different models employed. Subsection 3.2 presents a hyperparameter optimization study to determine the optimal values for the discount factor ($\beta$) and the size of the prediction window ($T_W$). Subsection 3.3 compares the performance of the different models considered. Subsection 3.4 compares the behavior of the models with that of classical two-stage programs and programs based on perfect information. Finally, subsection 3.5 compares the performance of the models considered with a single-phaseequivalent of the three-phase unbalanced MG.

*3.1. Case study*

In this work, the modified IEEE 34-bus test system [11] is utilized to simulate the behavior of a grid-connected MG. The system incorporates five solar farms, two wind turbines, two diesel generators and one battery system, as depicted in Figure 6. The main parameters of the MG and the devices installed in it are shown in Table 2. The data for the hour-by-hour variation of energy prices were obtained from [40]. Data for solar generation, wind generation and active and reactive power demand were taken from [11].

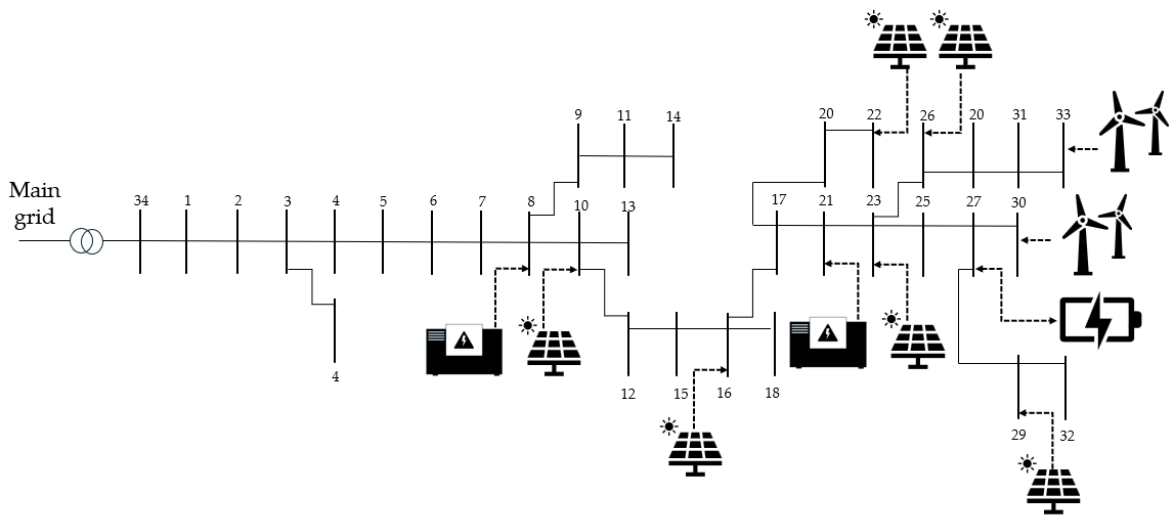

*Figure 6.Modified IEEE-34 bus test system.*

Following [11], error in forecasted variables is modeled as a normal probability distribution function, $N(\mu, \sigma^2)$ for each hour in the considered time horizon [11]. Thus, to generate the true realization of stochastic variables $\vec{W_t}$ (see Figure 5), values for each variable are sampled from these normal distributions for each time step, as done in [11].

Simulations were conducted over a time horizon of 24 hours, with timesteps of 1 hour, using a 32-cores, 128 GB of RAM Intel Xeon ® computer running a Linux based operating system. Both the model algorithms and the environment were implemented using the off-the-shelf software Gurobi 11, and each model is allocated a maximum solver runtime of one hour.

| No. | Parameter | Value |
| --- | --- | --- |
| 1 | Solar farms capacity | 300kW |
| 2 | Wind turbines capacity | 600kW HW43 |
| 3 | Battery's capacity | 3.9 MWh Tesla Megapack |
| 4 | Diesel generator's capacity | 800 kW Perkins |
| 5 | Nominal voltage | 24.9 kV |
| 6 | Nominal apparent power | 100 kVA |
| 7 | Diesel price | 4.23 kr/kWh |
| 8 | Maximum allowed voltage | 1.05 p.u |
| 9 | Minimum allowed voltage | 0.95 p.u |
| 10 | Maximum allowed electrical current | 46 A |
| 11 | Power factor for DERs | 0.95 |

*Table 2. Main parameters of the MG.*

Table 3 outlines the five models explored in this study: Quadratic nonconvex, Quadratic convex, and three linear approximations based on polygons of 8, 16, and 32 sides.

*3.2. Hyperparameter optimization*

As indicated in Section 2.1 and 2.3, the proposed models depend on two hyperparameters: the prediction window size $T_W$ and the discount factor $\beta$. To determine appropriate values for these hyperparameters, the open-source library Optuna [33] was used. A total of 25 trials were conducted -- each comprising 100 simulations -- using the Linear 2-sided polygon model (see Table 3) as the reference. The ranges for $T_W$ spanned 1 to 24, whereas $\beta$.varied from 0 to 1.

As shown in Figure 7, the best objective values are obtained with a $T_W$ of 21 hours and a discount factor close to 1. Specifically, the Optuna study identified 21 hours and $\beta \approx 0.997$ as the optimal values. However, as illustrated in Figure 7, no substantial improvement is observed for window sizes exceeding 11 hours. A similar trend was noted across all models considered in this work. Therefore, in the remainder of this paper a window size of 11 hours and a discount factor of 0.997 were used.

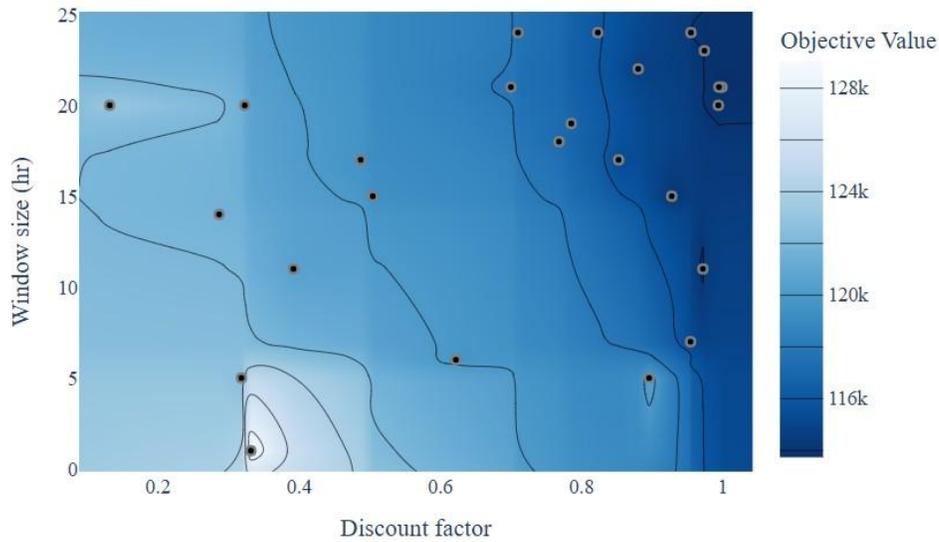

*Figure 7. Optuna study's results. Each red point corresponds to a trial with 100 simulations.*

*3.3 Comparison of model performance*

To thoroughly evaluate the five models listed in Table 4, 1000 simulations were carried out for each model. All forecasts and "true" realizations of the stochastic variables (see Figure 3) were held constant across these runs, ensuring consistency in the comparison. The prediction window size and discount factor were fixed at 11 hours and $\beta=0.997$, respectively, based on the hyperparameter exploration described above.

| Model name | Description |
|---|---|
| QCP nonconvex | Objective function (*4*) together with constraints (*5*) – (*40*) with the linearization process described in section 2.3.1 |
| QCP convex | The same as quadratic nonconvex but with the linearization process described in section 2.3.2 |
| Linear 2-sided polygon | The model described in section 2.3.3 using a 2-sided polygon in the first quadrant to linearize quadratic constraints |
| Linear 4-sided polygon | The model described in section 2.3.3 using a 4-sided polygon in the first quadrant to linearize quadratic constraints |

*Table 3. Models proposed in this work.*

Figure 8 compares the performance of the different models. Notably, the QCP convex model achieves the highest average performance. Moreover, linear models exhibit a performance comparable to that of the QCP convex model when four-sided polygons are used to linearize the quadratic constraints. In contrast, the QCP non-convex model demonstrates the worst performance, as the off-the-shelf software fails to reach optimality within the allocated computation time. A wall time of 10 minutes was imposed, given that the EMS must decide at the beginning of each hour.

*3.4 Comparison with perfect information and two stage programs*

To assess how forecast uncertainty impacts decision-making quality, each model was additionally evaluated assuming perfect information. Figure 9 shows that the QCP nonconvex model with perfect forecasts marginally outperforms the same model under uncertain forecasts, with a mean difference of approximately 1.48%. This difference can be interpreted as the upper bound on the benefit of improving forecast accuracy.

Figure 10 also contrasts the proposed models against a myopic model and a two-stage stochastic program based on [11], [24], [26]. In the two-stage program, all decisions for the 24-hour horizon are defined at $t = 0$ using a fixed set of scenarios, with no opportunity to re-optimize as actual conditions unfold. In the context of this work, a myopic model is defined as one that has no access to any forecast, allowing it to compute an implementable decision for the current time stage, without considering information about the future. Although two-stage approaches perform slightly better than the blind model, they are consistently outperformed by the proposed models, which dynamically update decisions as new information becomes available.

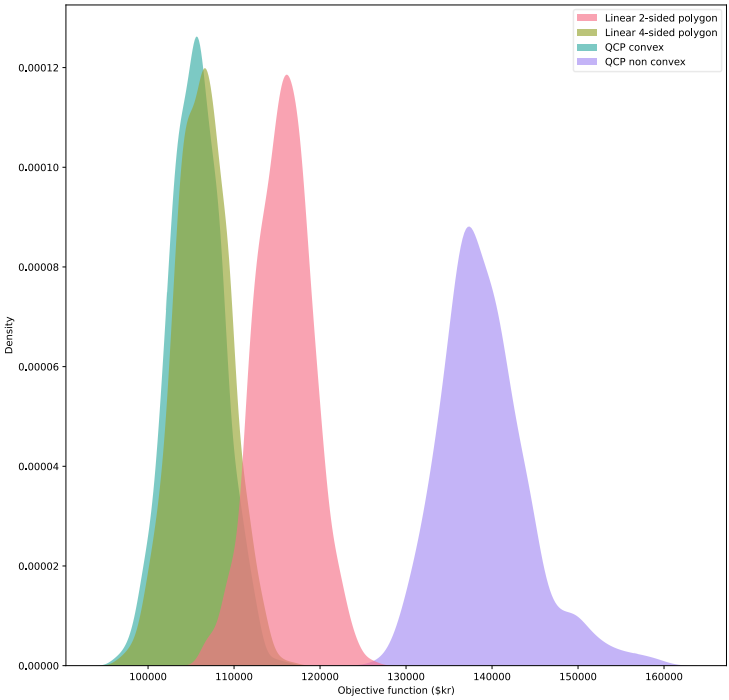

*Figure 8. Comparison of models' performance (see Table 3)*

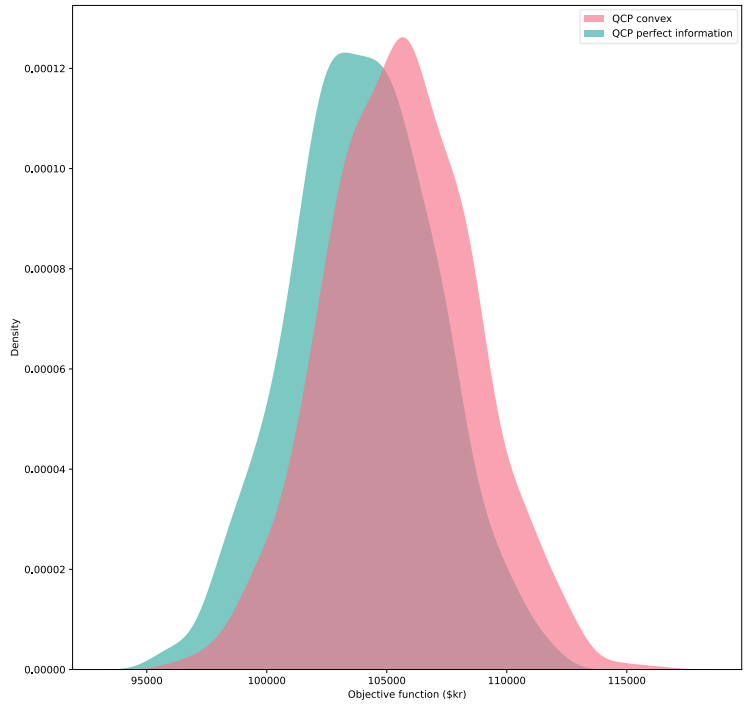

*Figure 9. Comparison of QCP convex model with and without perfect information.*

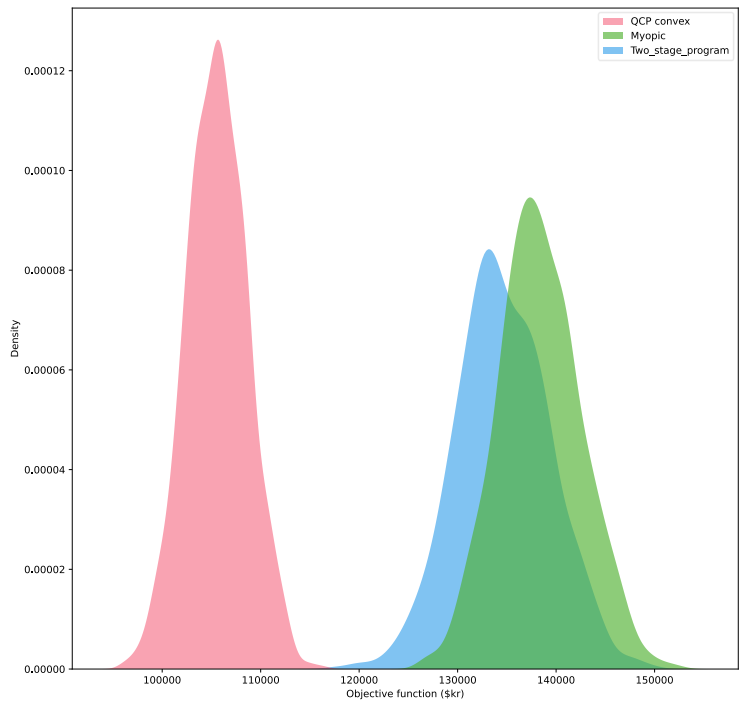

*Figure 10. Comparison of QCP convex behavior against the Two-Stage program and Myopic approach*

*3.5 Comparison with a single-phase equivalent*

To compare the performance of the unbalanced three-phase MG model against a model based on a single-phase equivalent, the model based on a single-phase equivalent [11] was implemented. shows the performance of the Linear 2-sided polygon model compared to the model based on the single-phase equivalent with the same linearization scheme. It can be seen that unbalanced three-phase model performs significantly better. Specifically, for the set of simulations used, the expected performance of the unbalanced three-phase model is 7.5% higher than that of the single-phase equivalent.

*3.6 Computing time*

The computational time of the QCP non-convex, linear with two-sided polygons, and linear with four-sided polygons models is compared in Figure 12. As observed, the computation times for all three models remain consistently below 2 seconds. This result demonstrates the applicability of these convexified and linearized models for real-time decision-making, where a rapid EMS solution is required.

*3.7 Voltage behavior*

In Figure 13, Figure 14, Figure 15, the behavior of voltage ranges for the three phases of the three-phase system over 24 hours is depicted. It can be observed that the effect of solar power injection within the time window between 8 AM and 4 PM results in an increase in voltage ranges above 1 p.u. Additionally, it is evident that the EMS successfully maintains voltage values within the defined operational limits.

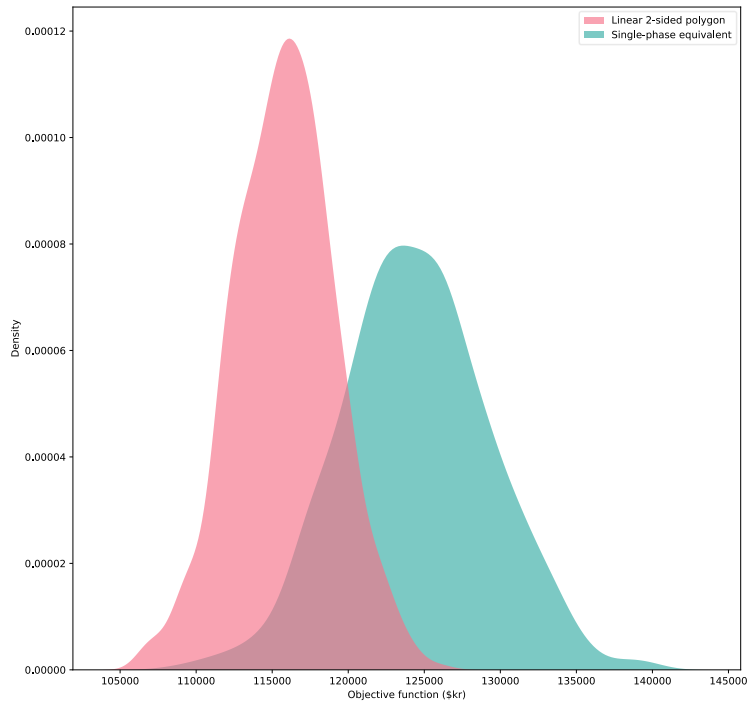

*Figure 11. Behavior of QCP convex model against a model based on a single-phase equivalent.*

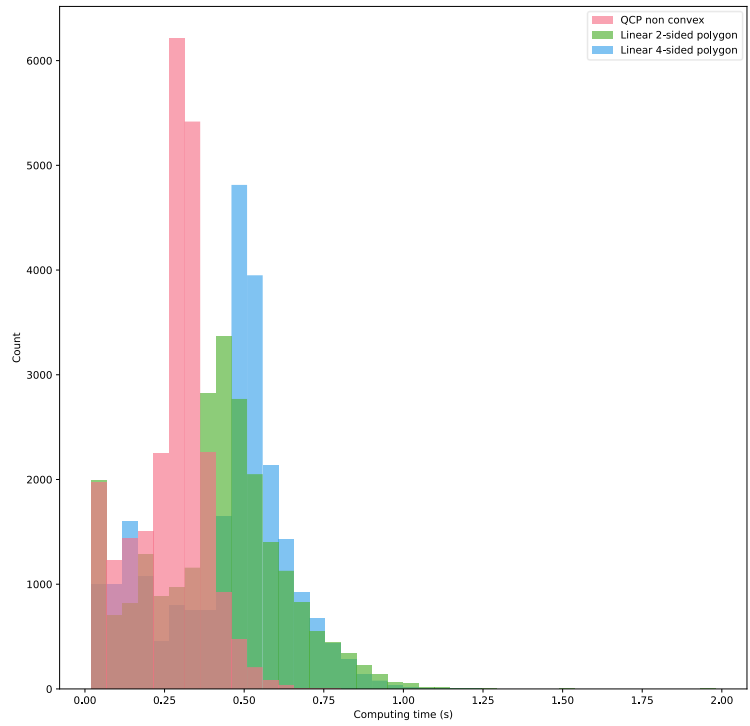

*Figure 12. Computing times for QCP nonconvex, linear 2-sided polygons and 4 -sided polygons.*

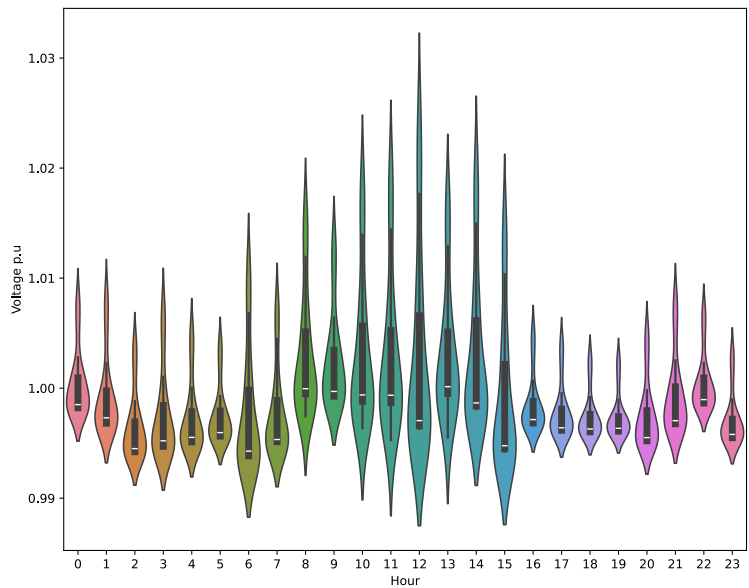

*Figure 13. Voltage behavior for phase a*

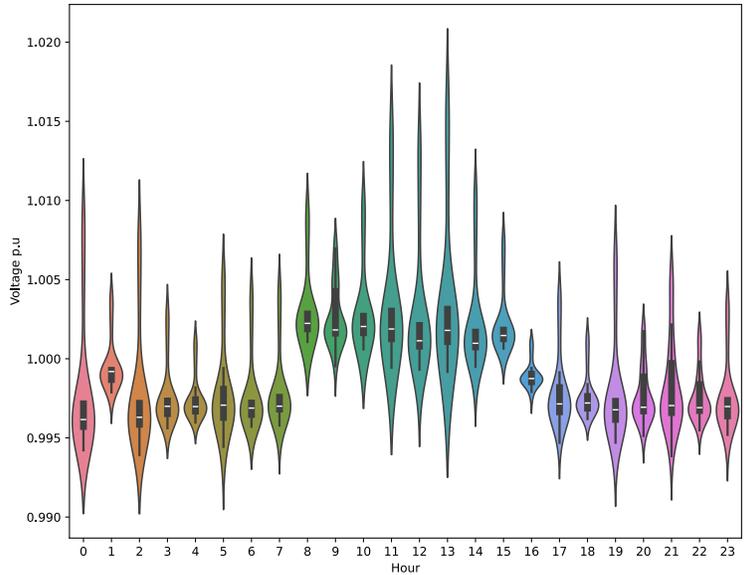

*Figure 14. Voltage behavior for phase b*

However, in extreme scenarios, due to the linearizations and convexifications applied, voltages may deviate beyond the feasible region. Among the 1000 simulations performed, only 0.3% of the cases exhibited voltage values outside the feasible region. Modifying these actions to ensure their safety could be an interesting research avenue for future developments.

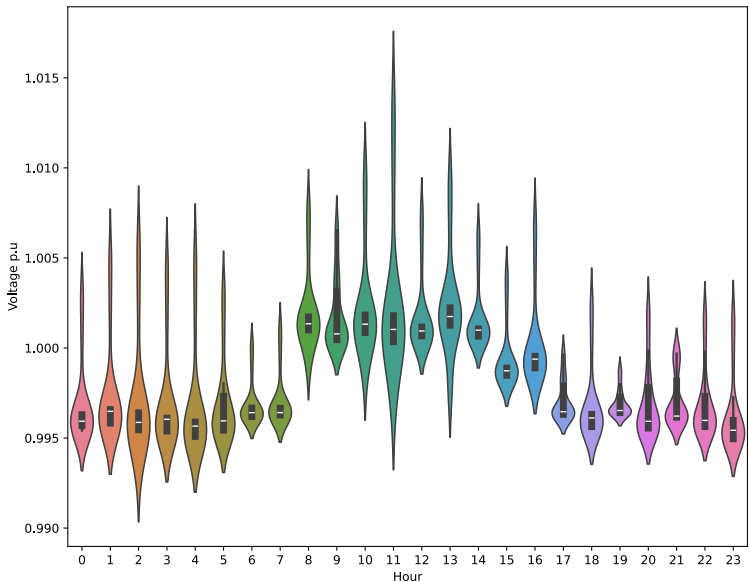

*Figure 15. Voltage behavior for phase c*

## 4. Conclusions

In this work, a linearization technique for the energy management problem in three-phase unbalanced MGs, current-based, was proposed. The performance of quadratic convex, quadratic non-convex, and linear models with different degrees of refinement were tested under various conditions. The results demonstrate that linear models, when sufficiently refined with a greater number of sides in the polygons used to approximate the quadratic constraints, can perform nearly as well as the quadratic non-convex model. In contrast, the quadratic nonconvex model exhibits the worst performance, primarily due to the limitations of off-the-shelf software that was unable to achieve optimality, thus forcing the model to take suboptimal actions. This highlights the ongoing challenge of solving quadratic non-convex problems and non-linear cases within computing times of less than one hour, as noted by [19]. Furthermore, the linear model proposed in this study offers a significant advantage by enabling implementation using widely available, open-source software. This is especially important as MGs become more prevalent, particularly in underserved communities seeking access to advanced energy technologies. Additionally, the linear model provides a foundation for further research into alternative solution methods, such as Benders decomposition [31], [41], Nested Benders decomposition [31], SDDP [32] among others. Moreover, the fact that both the linear model and the convex QCP can be solved by off-the-shelf software in less than two seconds is a promising result for their real-time implementation in microgrids, in contrast to computationally expensive models such as that of [15].

## 5. Future work

Future research in this field could focus on addressing the energy management problem in three-phase unbalanced MGs through the application of quantum computing. This emerging technology has the potential to solve particularly complex optimization problems in a fraction of the time, potentially revolutionizing the efficiency of such systems. Another promising avenue for future work is the exploration of Benders decomposition, Nested Benders decomposition, and Stochastic Dual Dynamic Programming (SDDP) methods for two-stage stochastic models. These techniques may offer more efficient solutions for managing the uncertainties inherent in MG operations. Finally, the growing popularity of reinforcement learning techniques presents an exciting opportunity to investigate their applicability to the energy management problem in three-phase unbalanced MGs. Leveraging reinforcement learning could lead to more adaptive and autonomous systems, further enhancing the performance and scalability of MGs in real-world applications.

APPENDIX A:

The corresponding terms for the linearization of expressions

$$I_{n,f}^{Dre} = \frac{P_D * V_{re} + Q_D * V_{im}}{V_{re}^2 + V_{im}^2} = g(P_D, Q_D, V_{re}, V_{im})$$

$$I_{n,f}^{Dim} = \frac{P_D * V_{im} - Q_D * V_{re}}{V_{re}^2 + V_{im}^2} = h(P_D, Q_D, V_{re}, V_{im})$$

may be derived by applying Taylor expansion as follows:

$$I_{n,f}^{Dre} \sim g^* + \frac{\partial g}{\partial V_{re}}_* (V_{re} - V_{re}^*) + \frac{\partial g}{\partial V_{im}}_* (V_{im} - V_{im}^*)$$

$$I_{n,f}^{Dim} \sim h^* + \frac{\partial h}{\partial V_{re}}_* (V_{re} - V_{re}^*) + \frac{\partial h}{\partial V_{im}}_* (V_{im} - V_{im}^*)$$

Where:

$$\frac{\partial g}{\partial V_{re}} = \frac{P_D * (V_{re}^2 + V_{im}^2) - 2 * Q_D * V_{re} * V_{im}}{(V_{re}^2 + V_{im}^2)^2}$$

$$\frac{\partial g}{\partial V_{im}} = \frac{Q(V_{re}^2 - V_{im}^2) - 2P_D * V_{re} * V_{im}}{(V_{re}^2 + V_{im}^2)^2}$$

$$\frac{\partial h}{\partial V_{re}} = \frac{Q_D * (V_{re}^2 - V_{im}^2) - 2P_D * V_{re} * V_{im}}{(V_{re}^2 + V_{im}^2)^2}$$

$$\frac{\partial h}{\partial V_{im}} = \frac{P_D * (V_{re}^2 - V_{im}^2) + 2Q_D V_{re} V_{im}}{(V_{re}^2 + V_{im}^2)^2}$$

In the case of the distributed energy sources (DER), the linear expression are extended to:

$$I_{n,f}^{Dre} \sim g^* + \frac{\partial g}{\partial V_{re}}_* (V_{re} - V_{re}^*) + \frac{\partial g}{\partial V_{im}}_* (V_{im} - V_{im}^*) + \frac{\partial g}{\partial P}_* * (P - P^*) + \frac{\partial g}{\partial Q}_* (Q - Q^*)$$

$$I_{n,f}^{Dim} \sim h^* + \frac{\partial h}{\partial V_{re}}_* (V_{re} - V_{re}^*) + \frac{\partial h}{\partial V_{im}}_* (V_{im} - V_{im}^*) + \frac{\partial h}{\partial P}_* * (P - P^*) + \frac{\partial h}{\partial Q}_* (Q - Q^*)$$

Where:

$$\frac{\partial g}{\partial P} = \frac{V_{re}}{V_{re}^2 + V_{im}^2}$$

$$\frac{\partial g}{\partial Q} = \frac{V_{im}}{V_{re}^2 + V_{im}^2}$$

$$\frac{\partial h}{\partial P} = \frac{V_{im}}{V_{re}^2 + V_{im}^2}$$

$$\frac{\partial h}{\partial Q} = -\frac{V_{re}}{V_{re}^2 + V_{im}^2}$$